\documentclass[showpacs,amsmath,twocolumn,showkeys,10pt,floatfix,aps,pra,longbibliography,floatfix]{revtex4-2}
\usepackage[T1]{fontenc}
\usepackage{inputenc}
\usepackage[margin=1in]{geometry}
\usepackage{color,times}
\usepackage{array}
\usepackage{amsmath}
\usepackage{stackrel}
\usepackage{graphicx}
\usepackage{appendix}
\usepackage{mathrsfs}
\usepackage{amsfonts}
\usepackage{appendix}
\usepackage{hyperref}
\usepackage{color}
\usepackage{float}
\definecolor{med-blue}{RGB}{25,25,112} 
\hypersetup{colorlinks, linkcolor={red},citecolor={blue}, urlcolor={blue}}
\allowdisplaybreaks
\DeclareMathOperator{\taninv}{\tan^{-1}}
\newcommand{\ket}[1]{\vert{#1}\rangle}
\newcommand{\bra}[1]{\langle{#1}\vert}

\newcommand{\expv}[1]{\langle{#1}\rangle}

\newcommand{\norm}[1]{\vert{#1}\vert^2}

\begin{document}

\title{An efficient quantum-classical hybrid algorithm for distorted alphanumeric character identification}
\author{Ankur Pal$^{1}$, Abhishek Shukla$^{2,3}$ Anirban Pathak$^{4}$}
\affiliation{$^{1}$Indian Institute of Science Education and Research Kolkata, Mohanpur, Nadia 741246, West Bengal, India}
\affiliation{$^{2}$ IMOMEC division, IMEC, Wetenschapspark 1, B-3590 Diepenbeek, Belgium}

\affiliation{$^{3}$Institute for Materials Research (IMO), Hasselt University, Wetenschapspark 1, B-3590
Diepenbeek, Belgium}
\affiliation{$^{4}$ Department of Physics and Materials Science \& Engineering\\ Jaypee Institute of Information Technology, A-10, Sector-62, Noida, 201309, Uttar Pradesh, India}

\begin{abstract}
   An algorithm for image processing is proposed. The proposed algorithm, which can be viewed as a quantum-classical hybrid algorithm, can transform a low-resolution bitonal image of a character from the set of alphanumeric characters (A-Z, 0-9) into a high-resolution image. The quantum part of the proposed algorithm fruitfully utilizes a variant of Grover's search algorithm, known as the fixed point search algorithm. Further, the quantum part of the algorithm is simulated using CQASM and the advantage of the algorithm is established through the complexity analysis. Additional analysis has also revealed that this scheme for optical character recognition (OCR) leads to high confidence value and generally works in a more efficient manner compared to the existing classical, quantum, and hybrid algorithms for a similar task. 
\end{abstract}
\maketitle

\section{Introduction}\label{Intro}
Ever since Richard P. Feynman’s 1982 \cite{Feynman1982} paper on the viability of using quantum resources (computers) to simulate quantum mechanical systems appeared, researchers have proposed various ways in which we can harness the power of quantum mechanics to solve problems that are difficult to be solved using classical computers in a reasonable time. Interestingly, many of the problems that quantum computers can solve faster than their classical counterparts are apparently unrelated to quantum mechanics, or even physics. For example, the Bernstein-Vazirani algorithm \cite{bernstein1997quantum} and the Deutsch–Jozsa algorithm \cite{deutsch1992rapid} were solutions to problems that were unrelated to quantum mechanics. The problems that they solved were of proof of concept nature, having very little practical uses. Interestingly, there exists a set of quantum algorithms having important practical applications. These algorithms also perform relevant computational tasks faster than their classical counterparts. An excellent example is Shor’s algorithm \cite{shor1994algorithms}, which has significant practical applications in the area of cryptanalysis. It essentially provides an efficient scheme for the factorization of the odd bi-primes, a problem not directly related to Physics. Here, we specifically stress on this aspect as we plan to propose an algorithm that will use quantum resources to solve a problem related to optical character recognition (OCR) which is also apparently unrelated to quantum mechanics. This just illustrates the large domain of applicability of quantum computing and does not indicate that quantum computing is not useful in solving problems closely connected to physics. Relevance of quantum algorithms in solving problems related to quantum physics is demonstrated through simulation of many-body Fermi systems on a universal quantum computer \cite{abrams1997simulation} and more recently through the realization of  a variational quantum eigenvalue solver using photonic quantum processors \cite{peruzzo2014variational,mcclean2016theory}.

 With time, it's realized that quantum computers are far more versatile than originally expected, and several new applications of quantum computing have been proposed. One such application is  Quantum digital image processing (QDIP)  which has recently been developed as a field that uses the power of quantum computation to solve a specific class of problems i.e., image processing-related problems which have
 applications in daily life. To understand its importance, we first need to note that a digital image is basically a digital object containing various visual information of an object of interest distributed over the space discretely and represented by a matrix. The smallest addressable element of the matrix is known as a pixel. The information about the color shade at every pixel is described by the pixel value (usually it's a number between 0-255 as an 8-bit/1-byte register is used for describing 256 shades of a bi-tonal image. Higher number of shades would require relatively large bit string, but that's not our concern at the moment). The full information of the shade of particular color is usually described by the parameters: hue, saturation, and luminosity \cite{gonzalez2009digital}. As long as any processing operation is linear and reversible and the related information can be encoded in the quantum systems, the traditional (classical) processing can be replaced with the unitary operation and a quantum algorithm can be used for obtaining the computational advantage \cite{yuan2013quantum}. As in this procedure, the basic process of image processing not only involves quantum resources, but also a quantum algorithm, it's referred to as QDIP.

 Before we describe our quantum-classical hybrid protocol for image processing, it would be apt to note that in the recent past a set of schemes for representing the image information using quantum systems (here termed as quantum digital image representation) have been developed and several applications of those have been proposed (for a review see  \cite{yan2016survey}). For example, the qubit lattice model is used to map the spatial information about the image into probability amplitudes of qubit states \cite{venegas2003storing}. Further, in Ref. \cite{le2011flexible}, a scheme is proposed that can not only encode the shade and color information, but also the spatial information i.e., pixel index. Similarly, quite a few quantum algorithms for edge detection have been proposed, like QSobel \cite{zhang2015qsobel}, Laplacian and Sobel algorithm based scheme proposed in\cite{fan2019quantum}, and improved Sobel operator based algorithm of Ma et al. \cite{ma2020demonstration}. Following an independent path, quantum support vector machine is deveopled and is used for the grouping of images for classification requiring feature map \cite{park2020practical}. Furthermore, a convolutional neural network has shown very good performance for exploiting the correlation information when the convolutional layer and pooling layer are replaced by the quantum layer \cite{cong2019quantum}. Schemes for QDIP usually improve efficiency and results of processing classically stored digital images, using protocols allowing us to interpret the measurement outcomes of qubits to represent classical information of the processed image. QDIP has many facets and a few algorithms for QDIP have already been designed for a set of specific image-processing tasks. Specifically, we may mention that algorithms have been designed for quantum image scrambling \cite{jiang2014The,jiang2014quantum,zhou2015quantum,arbelaez2010contour},  geometrical transformation of quantum image  \cite{le2010fast,le2011strategies}, quantum image scaling \cite{zhou2017quantum}, quantum image encryption and decryption \cite{zhou2013quantum}, quantum image steganography \cite{jiang2016lsb}, watermarking of quantum image \cite{iliyasu2012watermarking,yan2015duple}, quantum audio \cite{yan2018flexible}, quantum movie \cite{iliyasu2011framework}, quantum image segmentation \cite{caraiman2015image}, and quantum image matching \cite{jiang2016quantum}. Moreover, Tseng and Huang \cite{tseng2003quantum} has proposed an edge detection scheme that has the same performance as the Sobel method, while the Sobel method was proved better than the corresponding classical methods for detecting noisy images \cite{xie2005research}. In this particular work, we plan to propose a QDIP scheme that will be useful in image processing tasks of a specific type. Specifically, the quantum part of our hybrid algorithm would perfrom the pre-processing tasks of the overall OCR process for low-resolution printed text images. 
The motivation behind  the designing of QDIP scheme for OCR is manifold. Firstly, OCR is commonly used in our daily life. Secondly, OCR-related studies form one of the most active fields in machine learning (ML), quantum ML (QML), and quantum-assisted ML (QAML). The recent success of ML, QML and QAML algorithms in performing various tasks in an efficient manner hint at the possibility of designing an efficient hybrid algorithm for image processing related to OCR where quantum resources and the concepts QML and QAML can be used in a beneficial manner.

Nowadays, with the increase in computational power, we can handle datasets of enormous sizes, and that allows OCR software to recognize not only printed character text, but also handwritten texts \cite{deselaers2011latent}. However, efficiency and accuracy are still a concern and there still remains scope for improvement as far as the results for the aforementioned case of using OCR to identify printed text in a particular font are concerned. While training an OCR, we train OCR, for a particular type of font set. Every character belonging to that front set has a definite pixel matrix, we train OCR first to find which pixel matrix corresponds to that particular character. This ensures that we only need to provide a well-labeled dataset for each character and then rely on ML or QML  for theidentification of the character. So, when we input an image into an OCR system (image acquisition), it can check which character it resembles the most \cite{awel2019review}. This is a simplistic way of understanding how OCRs work, but on a very basic level, it interprets the information conveyed in images by the values representing each pixel. Typically, the overall process of OCR involves pre-processing consisting of normalization, denoising, and skew correction, followed by segmentation, feature extraction, and identification \cite{modi2017review}. 
These processes help in increasing the accuracy and efficiency of OCR algorithms. Keeping the importance of pre-processing in mind, in the following section we will describe the important prepossessing techniques that should be used before implementing our algorithm which will be described in the subsequent sections.

The rest of the paper is organized as follows. In Sec. \ref{pre-proc}, we briefly introduce the pre-processing techniques used in OCR algorithms as those will be required for our algorithm, too. Subsequently, in Sec. \ref{CharId}, we formulate the problem, introduce the quantum systems for encoding the relevant information, and describe the flow of the overall algorithm. In Sec. \ref{classical}, we describe the use of the classical part of the algorithm for pre-processing and for identification of the missing information and hence the right character. In Sec. \ref{quantumPart}, we describe the quantum part of the algorithm which is based on a fixed point search algorithm. In Sec. \ref{Results}, we present the confidence value obtained using our algorithm and analyze the results ending in a conclusion that our algorithm correctly and efficiently identifies the alphanumeric characters.
\section{Pre-processing techniques}\label{pre-proc}
One important step of OCR is pre-processing. This step is important because the input image itself may not be an ideal input for the OCR algorithm. Pre-processing is done to enhance the performance of OCRs. Some of the most commonly used pre-processing methods are$\colon$

\begin{itemize}
    \item \textbf{Image thresholding}: Image thresholding is the process of removing the background, for a bitonal image. It is done by setting the pixel value in a pixel matrix to 0 (1) if it is below (above) the threshold. It reduces the amount of data by removing irrelevant details from an image (see example images  for local thresholding  illustrated in \cite{wikiImageThresholding}). This method is useful when the image has been taken under irregular lighting. After removing the background, the pixel matrix (the matrix of pixel values) is normalized. 
\item \textbf{Skew correction}$\colon$ It is the process of aligning the images. An example of it can be seen in Figure 1 of Ref.  \cite{shafii2015skew}, which shows histograms of a document for two different skew angles. The histogram is created by summing up all the black pixels in the horizontal direction. For a non-zero skew angle, the number of pixels for a row number is distributed in an immiscible manner, while for a zero skew angle the histogram is regular. 

\item \textbf{Denoising}$\colon$ The main sources of noise in digital images are due to irregular or low light acquired by 
capturing device, sensor temperature effects, and sometimes due to the error in the mathematical model. The main methods used for 
denoising are mean filters, median filters, and periodic noise filtering (see text images in \cite{ImageNoiseReduction} and 
methods for denoising them). These denoising methods are part of preprocessing process as these methods help us to  clean an image before the implementation of the main algorithm for image processing.
\item \textbf{Segmentation} It separates each character present in the text. It involves many steps like inverting the image to make the background black and foreground white to reduce space, conversion to RGB format to introduce colors other than black and white, and segmenting the word image determined by the sum of the pixel value of the foreground image to be 1 or 0 and ultimately removing the over-segmentation. Examples of all the steps are shown in Figure 3 of Ref. \cite{choudhary2013new}.
\end{itemize}
These pre-processing methods solve particular problems that arise in specific situations. For example, in the case of automatic number-plate recognition, one of the most common problems that arise is the low resolution of the image, usually because of the poor quality of the camera or because of the large distance of the number plate from the camera. For such cases, the use of pre-processing techniques to improve results for number-plate recognition from the low resolution images is an area of active research in classical computing. In automatic number plate recognition from low-quality videos, piece-wise gamma correction was suggested to ensure illumination invariance \cite{ajanthan2013automatic}. It was also suggested in Ref. \cite{ajanthan2013automatic} that a scheme for character segmentation based on information about character placement on the license number plate will be different worldwide. As part of pre-processing, we can also use the resolution enhancement technique to get accurate results in cases where we get no results at all, or worse, there is the chance of misidentification of characters. 
In the following, we propose a hybrid (quantum and classical) algorithm for the identification of distorted alphanumeric characters. The prerequisite for our algorithm is the High-resolution image of all alphanumeric characters for the Font type of interest. The pixel information is encoded in the quantum system and the quantum algorithm is used for finding pixel values that are near to the reference images. The classical OCR is then used to identify the right character by calculating the confidence value obtained from OCR. In particular, the fixed point search algorithm is used for finding the missing pixel values with the help of high-resolution reference images of all characters. So, every low-resolution image will now have the same number of images as the number of reference images with updated pixel values, for which confidence values using OCR are calculated.

\section{Distorted alphanumeric character identification} \label{CharId}
\begin{figure}
    \includegraphics[width=8cm, trim=3.5cm 3cm 3cm 0cm]{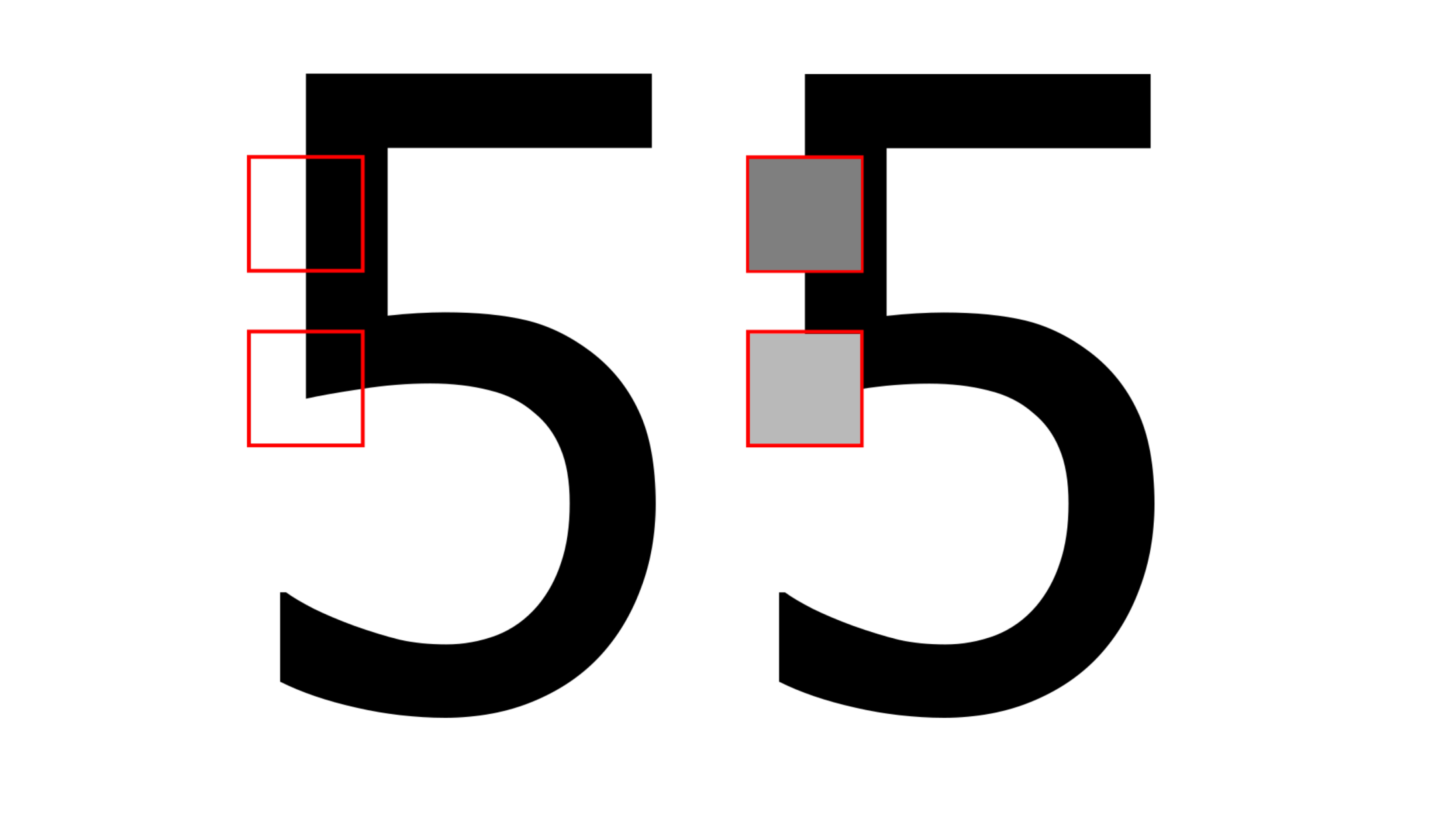}
    \caption{(Color online) Representation of assignment of pixel values when a digital camera captures an image.}\label{fig1}
\end{figure}

The smallest unit of accessible information in digital objects (an image or an alphanumeric text object) is known as a pixel. The 
resolution of an image is quantified by the number of pixels per inch (PPI), the greater the PPI the better is the resolution of the image. Every pixel contains information about bit depth, dynamic range, file size, compression, and metadata, which vary with the file format. The bit string corresponding to bit depth is used to store the information about the color and its shades as briefly described below.  Usually, the bit depth for the color images/text varies from 8 to 24 bits. Let's start with a grayscale image. For a grayscale image, a single bit is sufficient to encode black and white as 0 and 1, and the scale of grayness is defined by the pixel value (a value between 0 and 1) would signify
the particular shade of a grayscale image, in order to be able to represent different shades of a grayscale image multiple bits are required. For example, an 8-bit string can be used to store
$2^{8}$ shades of a single color, while for a color image represented under RGB primary color scheme total of 24 bits are required, 8 bits for each color. Thus, a 24-bit string can represent $\left(2^8\right)^3\sim16.7$ million colors. Though in this manuscript we just focus on finding missing information in a grayscale image of an alphanumeric character using a quantum bit (qubit) to store pixel value information. The quantum advantage in storing comes from the fact that a quantum bit can be prepared in infinitely many possible probability distributions and hence probability amplitudes can be used to store information about the shades i.e., pixel value. So one qubit is sufficient as a quantum resource that can store the necessary information for our algorithm.
Moreover, the use of a qubit may lead to a quantum advantage. Another advantage of our algorithm is it does not require a big dataset for training and is computationally less expensive unlike algorithms based on neural networks, e.g., in a recent work entitled, “A new image enhancement and super-resolution technique for license plate recognition” \cite{hamdi2021new}, a deep learning architecture has been used, which requires training on massive datasets and use of multiple loss functions, making the algorithm computationally expensive. So, we are only interested in the relative brightness of the pixel; say, 0 represents a black pixel, and 1 represents white. More specifically, when a digital camera assigns a pixel to an edge, it just takes the average value of all the values that lie within the confines of that pixel, as can be seen in Fig. \ref{fig1}. Here, the lower box will have a value closer to 1, and the upper box will have a lower value. In a grayscale image, the only information we have for that block is the average value. Hence, if there are very few pixels, much information needed by OCRs to identify the characters get lost. Therefore, we must find a way to retrieve that information from its average value only. This is where the classical part of our hybrid algorithm comes in. In the following section, we will first describe the classical part of the proposed hybrid algorithm. Subsequently, we will describe the quantum part of the algorithm in the next to next section.
\section{Classical Part of the algorithm}\label{classical}
The critical problem to solve is to find a way to retrieve the missing information needed to convert a low-resolution image to a high-resolution image. 
This can be made possible for character recognition of printed text, particularly if the typeface of the text is known because the 
missing information is restricted to a finite set (character set). To begin with, we can assume one by one, for each element of that set that it holds the missing information and try to process our image under that assumption. We can then assess which assumption is correct by assessing which of the processed image is best. Specifically, we compare the low-resolution image of the alphanumeric character whose resolution we want to increase with the high-resolution images of all alphanumeric characters of the same font through confidence value obtained by OCR. The alphanumeric character for which the confidence values (which is the measure of how confident are we about the character recognized by the OCR process) of the low-resolution image is closest to the confidence value of the high-resolution image is the identified character.    The confidence value is defined differently for different OCRs. To be specific, let us assume that we are dealing with an English alphanumeric character set (0-9, A-Z). Suppose that we get a low-resolution image of a character that our OCR cannot identify. If we assume that the missing information to convert that low-resolution image into a high-resolution image is in the character 0, we will get the processed image such that, if we run it through OCR, the confidence value for 0 will indicate how good our assumption is. We can similarly proceed with all the characters in the character set and choose the one with the highest confidence value. 

\begin{figure}[htbp]
    \includegraphics[width=8cm, trim=2cm 2cm 2cm 2cm]{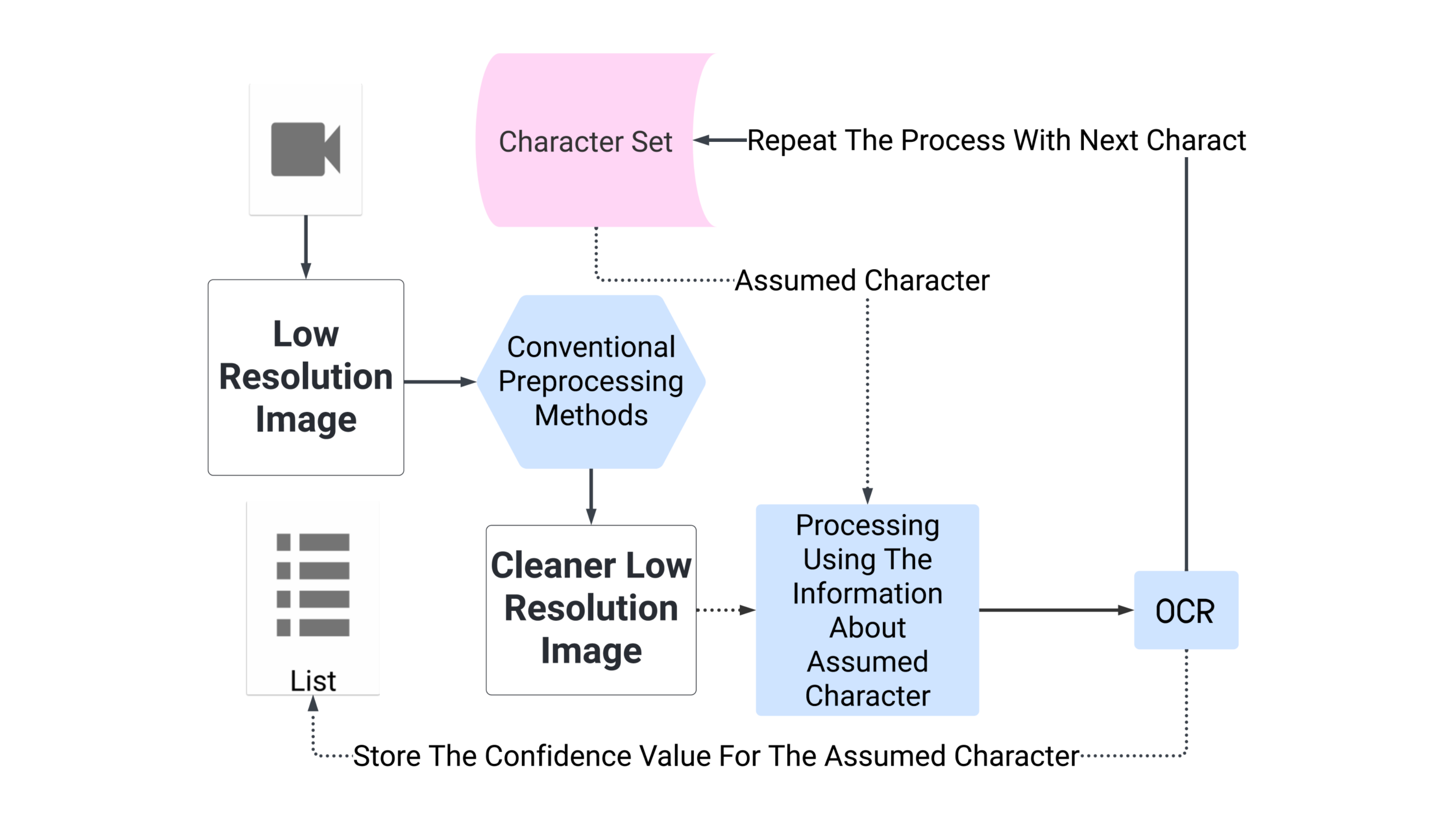}
    \caption{(Color online) In this figure, the conventional pre-processing methods
    include the applicable pre-processing methods mentioned earlier. After segmentation,
    we can treat each segment as a separate image. This image of a single character can
    then go through further processing using the information about an assumed character
    from the character set. After this process, we input this image into the OCR, from which
    we get the confidence value for the assumed character. We can store this value in a list. Then we repeat
    the process with the next character from the character set. The character which corresponds to the highest
    confidence value will be the answer. In this figure, the dotted lines represent the flow of information
    and the solid lines represent the sequence.}\label{fig2}
\end{figure}

A representation of this process can be seen in Fig. \ref{fig2}. Now, we will discuss the process through which we convert a low-resolution image into a high-resolution image using the information from the assumed character in quantum part of the algorithm.

\section{Quantum Part of the Algorithm} \label{quantumPart}
\begin{figure}[htbp]
    \includegraphics[width=8cm, trim=0cm 2cm 0cm 2cm]{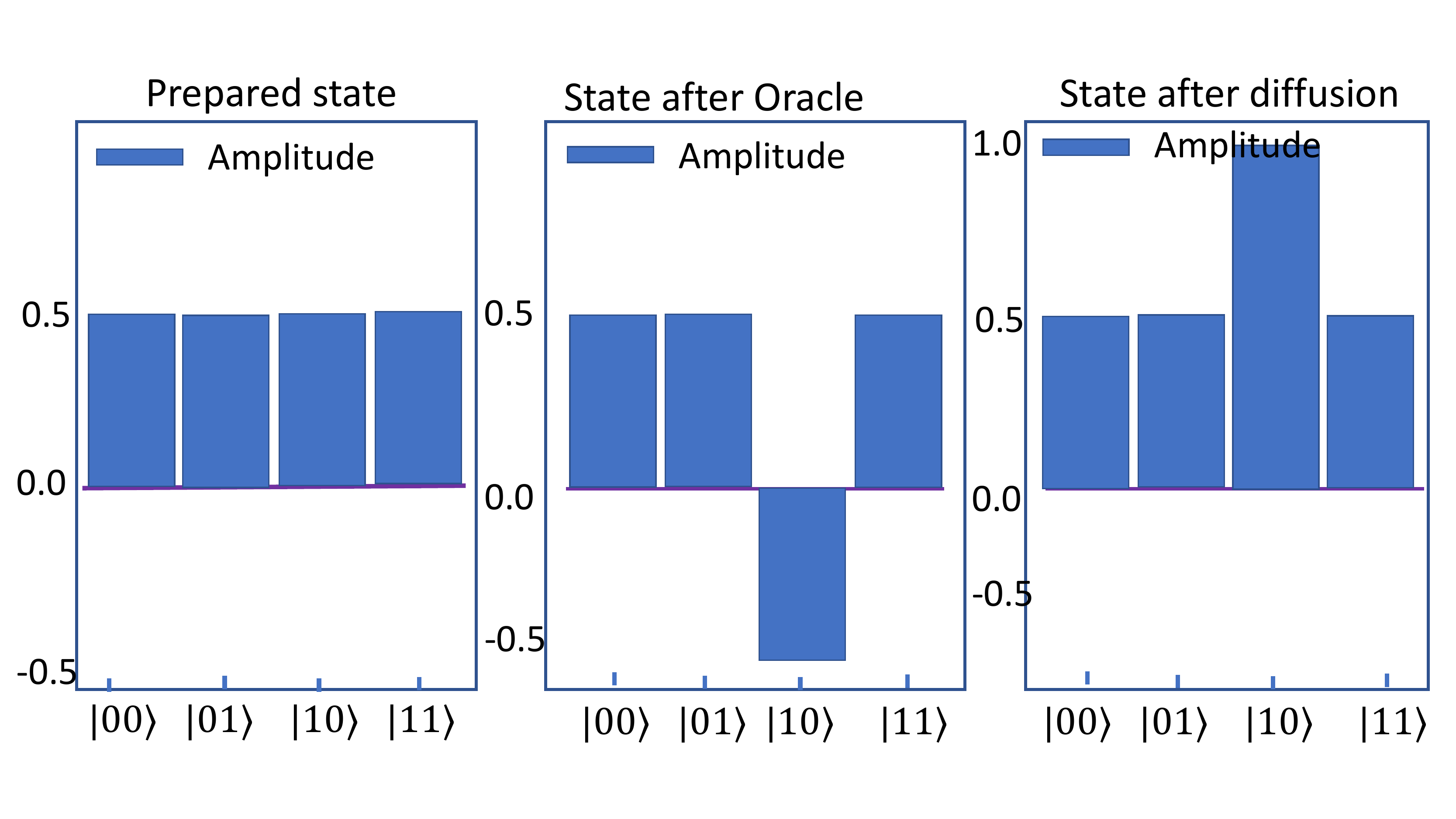}
    \caption{(Color online) This is an example of Grover’s search algorithm for a two-qubit system. Here, the search state is $\ket{10}$. After preparing an equal superposition of the states $\ket{00}$, $\ket{01}$, $\ket{10}$, and $\ket{11}$, the oracle reflection is applied to the state, resulting in rotation of $\pi$ of the search state. Finally, we apply the diffuser operation to get the search state.}\label{fig3Groversearch}
    \end{figure}

\begin{figure}[htbp]
    \includegraphics[width=8cm, trim=1cm 1cm 1cm 1cm]{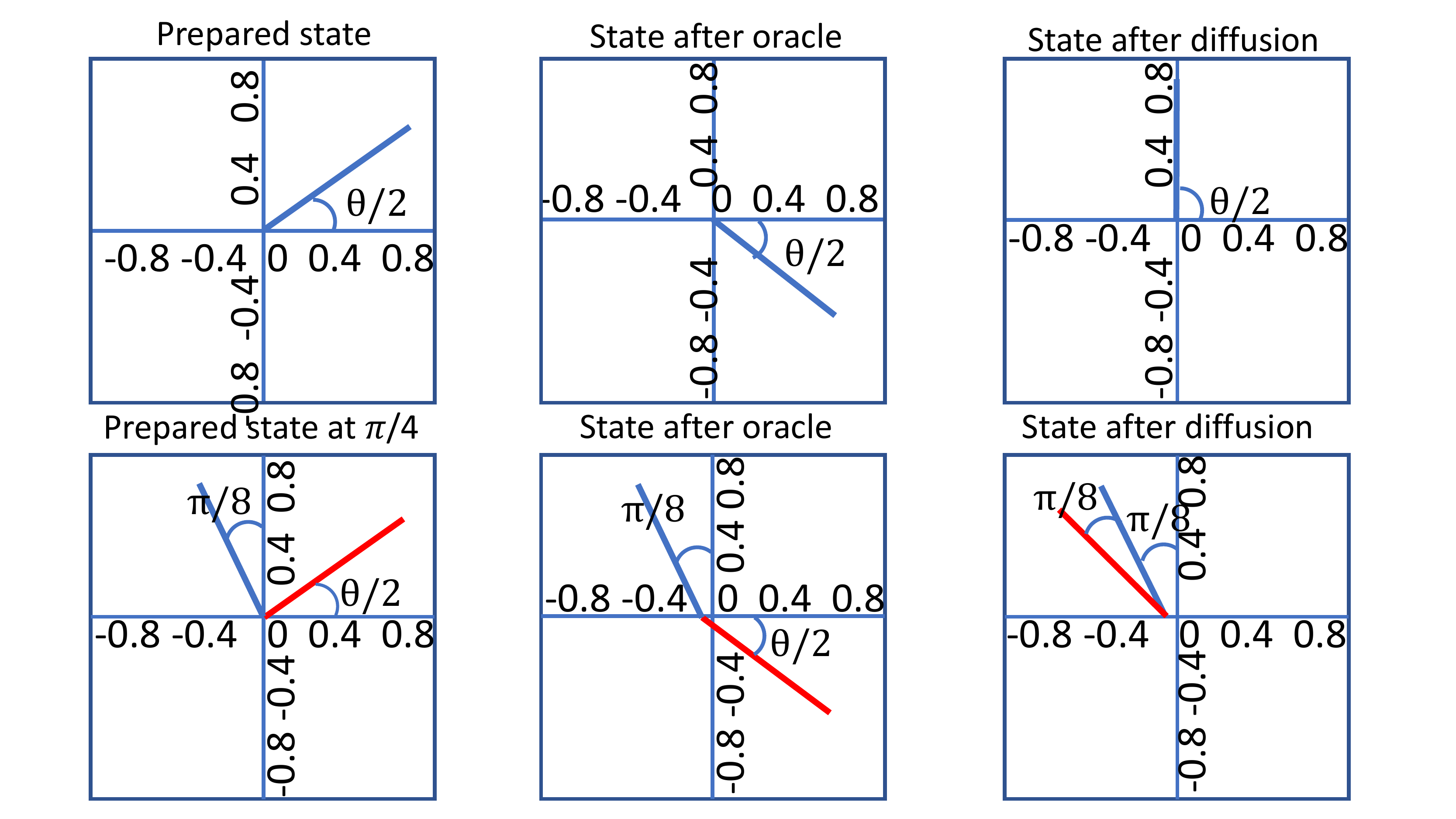}
    \caption{(Color online) These three sub-figures represent the same state in the corresponding sub-figures of Fig. \ref{fig3Groversearch}. Here, if we take the angle made by the initial state vector with the $x$-axis to be $\frac{\theta}{2}$, the oracle can be thought of as a reflection about the $x$-axis, and the diffuser can be thought of as reflection about
the initial state vector, resulting in the angle of the final state vector with the $x$-axis being 3$\frac{\theta}{2}$. When combined the result is the same as rotation by an angle of $\theta$. Taking the example of a target state such that it makes an angle of $22.5^{\rm{o}}$ with $\ket{0}$, for the initial state which makes an angle of $45^{\rm{o}}$ with the $x$-axis, the final state that we will get will not be the same as target state, however, it will also make an angle of 22.5$^{\rm{o}}$ with $\ket{0}$. Therefore, the probability of getting the state $\ket{0}$ on the measurement will be the same for both.}\label{fig4GroverSearchImplementation}
\end{figure}

\begin{figure}[htbp]
    \includegraphics[width=8cm,trim=0cm 0cm 0cm 0cm]{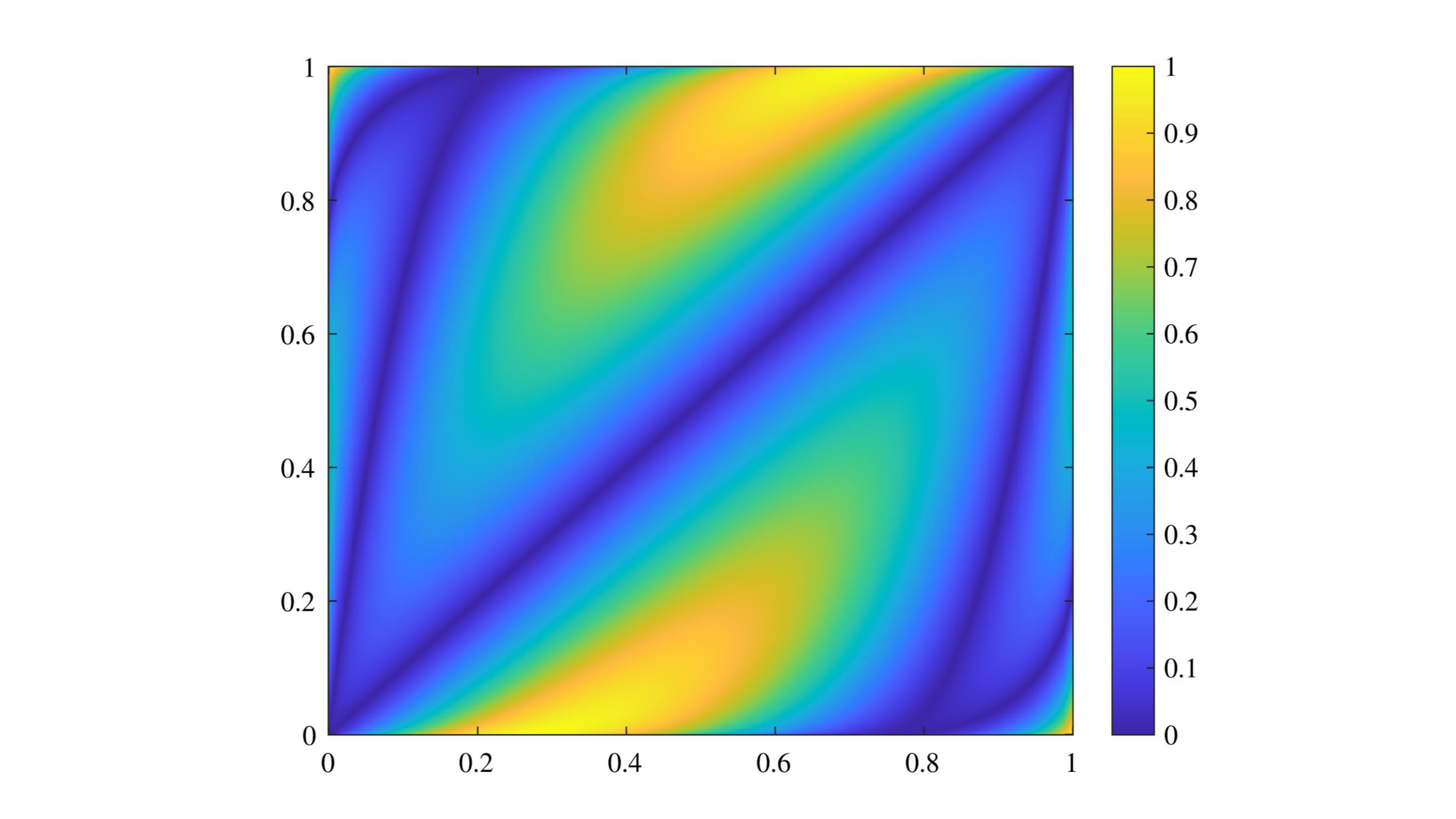}
    \caption{(Color online) Here, the $x$-axis is the probability of measuring the $\ket{0}$ state for the initial state $\ket{s}$ and the $y$-axis is the probability of measuring the $\ket{0}$ state for the target state $\ket{w}$. The color map
represents the difference between the probability of measuring the $\ket{0}$ state for the final state and the probability of measuring the $\ket{0}$ state for the target state $\ket{w}$. We can see that for states at $45^{\rm{o}}$
line ($\ket{s} = \ket{w}$), we get perfect matches as is expected since these states will have $\theta = \pi$ when represented in $\ket{w}$ and $\ket{s'}$ state. Then, we can see towards the top-left and bottom-right edges of
the figure, we have two curves, which also have perfect matches. These two curves represent the
states for which $\theta = \frac{\pi}{3}$ and $\theta = 5\frac{\pi}{3}$, respectively. There are also two curves that connect the
curve for which $\theta = \pi$ to the curves for which $\theta = \frac{\pi}{3}$ and $\theta = 5\frac{\pi}{3}$. For these two curves, $\theta \in (\frac{\pi}{3}, 5\frac{\pi}{3}$). These are the cases for which even though the amplitude for the final state and the target state $\ket{w}$ is different, the probability of measuring the $\ket{0}$ state will be the same, since they are at an equal angle from $\ket{0}$(See lower trace of Fig. \ref{fig4GroverSearchImplementation}). We can also see that for the case where $\ket{w} = \ket{0}$ or $\ket{w} = \ket{1}$, the difference between the probability of measuring the $\ket{0}$ state for the final state and the probability of measuring the $\ket{0}$ state for the target state $\ket{w}$ is 1 for two values. One of the cases is trivial as the initial state $\ket{s}$ is orthogonal to the target state $\ket{w}$ (top-left corner and the bottom-right corner).
However, for second case, the value of $\theta = 2\frac{\pi}{3}$, hence the final state will be orthogonal to the
target state $\ket{w}$.}\label{fig6Improvement}
\end{figure}

A qubit is the smallest unit of quantum information, which can be realized using a quantum system. A general two level quantum state $\ket{\psi}$ is a superposition of two eigenstates in the computational basis say, $\ket{0}$ and $\ket{1}$, where $\ket{\psi} = a\ket{0} + b\ket{1}$. Quantum algorithms for QDIP are composed of quantum operations on quantum systems designed for achieving enhancement in performance of the digital image processing. The benefits of such a quantum algorithm are multitudinous. 
For example, classically, if we represent a pixel in grayscale, we need 8 bits, assuming that we want 256 shades of gray (including black and white). However, we can represent one pixel in grayscale by using just a single qubit. Additionally, we can assign any value between 0 and 1 to a qubit, making it possible to represent infinite shades of gray. This is not advantageous if we have to access the pixel directly since the probabilistic nature of quantum mechanics or more precisely the collapse on measurement postulate of quantum mechanics implies  that we  have to prepare multiple qubits and measure them. We will still get a computational advantage.  In Fig. \ref{fig2}, it is mentioned that after getting a cleaner low-resolution image, we will process it using the information about the assumed character, which is the quantum part in our hybrid algorithm. In particular, we propose to use amplitude amplification as described in Grover algorithm \cite{grover1996fast}. Consider that There is an unsorted database containing $N$ items (say, $N$ alphanumeric characters) out of which just one item (character) satisfies a given condition and that need to be retrieved. The most efficient classical algorithm for this is to examine the items one by one requires about $\frac{N}{2}$ steps. While Grover algorithm takes $\sqrt{N}$ number of steps. We can use this algorithm for amplitude amplification of arbitrary input and search states, but that will not lead to the desired result because of the reasons described below. Ideally, we would require the difference in amplitudes of the final state and that of the search state of the low-resolution image to be directly proportional to the difference between the amplitudes of the final state and the search state of the high-resolution image. This would mean that if the low$-$resolution image has a pixel that corresponds to the average value of $n$ pixels of the high-resolution image we can take new $n$ pixels with the value that equals that of the low-resolution image, and subsequently, perform amplitude amplification on each of them with search state being that of the corresponding $n$ pixels of the high$-$resolution image. However, this is not possible if we use the generalized versions of the operations described in Grover’s original paper \cite{grover1996fast}. We can understand the problem by understanding Grover’s search algorithm by visualizing what each operator does. For this, we can have a look at the example shown in Fig. \ref{fig3Groversearch}. To generalize this, consider the search state as $\ket{w} = \ket{10}$, and consider the initial state as $\ket{s} =\frac{1}{2}(\ket{00} + \ket{01} + \ket{10} + \ket{11})$, where the oracle can be generalized as $O=\mathbb{I}-2\ket{w}\bra{w}$, and the diffuser as 
$D=2\ket{s}\bra{s}$. We can now write $\ket{s} = \frac{1}{2}\ket{w} + \frac{\sqrt{3}}{2}(\ket{00} + \ket{01} + \ket{11})$, and the $O$ and $D$ operators are defined such that the result spans the space defined by $\ket{w}$ and $\ket{s'}$, where $\ket{s'} = \frac{1}{\sqrt{3}}(\ket{00} + \ket{01} + \ket{11})$, in this particular case. If we define $\frac{\theta}{2}= \taninv \left ( \frac{\expv{s|w}}{\expv{s'|s}}\right)$, we can see that the entire operation is merely a rotation by an angle $\theta$. The 
representation of these transformations can be seen in Fig. \ref{fig4GroverSearchImplementation}. This, however, is problematic for us as rotation in the Grover algorithm can miss the target state.  In our low-resolution 
image, the value of the pixel can be anything between 0 and 1. If we now consider the case of a single qubit, the initial arbitrary state $\ket{\psi}$ can be written as a linear combination of search state $\ket{w}$ and its orthonormal state $\ket{s'}$. We can observe that for $\theta \in \{\frac{\pi}{3},\pi,5\frac{\pi}{3}\}$,

we will get the same state as search state $\ket{w}$ (ignoring the global phase). It is important to remember that although we are representing the initial states and the states after subsequent operations in the basis of search state $\ket{w}$ and its orthonormal state $\ket{s'}$, we measure in the computational basis. This means that for states for which $\theta \in \{\frac{\pi}{3},\pi,5\frac{\pi}{3}\}$, we can get an equal probability of obtaining $\ket{0}$ state if it is exactly in the middle of the final state and the search state $\ket{w}$. An example of this can be seen in the lower trace of Fig. \ref{fig4GroverSearchImplementation}. All of this will result in Fig. \ref{fig6Improvement}.  This is not suitable for our needs as   what we require is a perfect match for the case where the initial state and the target states are the same ($\ket{s} = \ket{w}$), and a monotonic increase in the difference between the probability of measuring the $\ket{0}$ state for the final state and the probability of measuring the $\ket{0}$ state for the target state $\ket{w}$. This is why Grover’s fixed-point quantum search algorithm is most suitable for our case \cite{grover2005fixed}. 


\subsection{Fixed-point quantum search algorithm} \label{FPSA}
The previously discussed algorithm can be generalized further by defining the initial state as $\ket{\psi} = U \ket{0}$,
oracle as $O_{\phi} = \mathbb{I}-(1-e^{i\phi})\ket{t}\bra{t}$, and the diffuser as $D_{\psi} = U(\mathbb{I}- (1-e^{i\phi})\ket{0}\bra{0})U^{\dagger}$. For $\psi = \phi = \pi$, we get the original algorithm \cite{grover2005fixed}. Let $||U_{ts}||^{2} = |\expv{t|s}|^{2} = 1 - \epsilon$, where $\epsilon$ is the deviation of the initial state $\ket{s}$ from the target state $\ket{w}$. If we now apply the operators $D_{\psi}O_{\phi}$ on $\ket{\psi}$, the component of the final state along $\ket{s'}$ will be $e^{\frac{i}{2}(\psi - \phi)}-4\sin(\frac{\psi}{2})\sin{\frac{\phi}{2}}|U_{ts}|^{2}$, which in the asymptotic limit $|U_{ts}| \rightarrow 1$, is minimized by taking $\psi = \phi = \frac{\pi}{3}$ \cite{tulsi2006new}. On applying the operators $D_{\frac{\pi}{3}}O_{\frac{\pi}{3}}$ on the initial state $\ket{s}$, we get the following superposition$\colon$
\begin{equation}
U\ket{s}[e^{i\frac{\pi}{3}} + ||U_{ts}||^{2}(e^{i\frac{\pi}{3}}-1)^{2}] + \ket{t}U_{ts}(e^{i\frac{\pi}{3}}-1).    
\end{equation}
On taking the square of the inner product of the above superposition with the orthonormal to the target state $\ket{s'}$,
we get the following$\colon$
\begin{equation}
    (1-||U_{ts}||^{2})||[e^{i\frac{\pi}{3}} + ||U_{ts}||^{2}(e^{i\frac{\pi}{3}}-1)2]||^{2} = \epsilon^3. 
\end{equation}
We can also use this algorithm recursively by taking the final state after one iteration as the initial state of the next iteration, such that for $m^{th}$ iteration$\colon$
\begin{equation}
U_{m} = U_{m-1}[\mathbb{I}-(1-e^{i\psi})\ket{0}\bra{0}]
U_{m-1}^{\dagger}O_{\phi}U_{m-1},
\end{equation}
where, $U_{0}\ket{0} = U\ket{0} = \ket{s}$.

This is the ideal result because if the pixels from the high-resolution image are taken as the target states and the pixels from the low-resolution images as the initial states, then the final image should resemble the high-resolution image if the high-resolution image is of the correct character.

\section{Results and Analysis} \label{Results}
Firstly, a low-resolution image of each alphanumeric character was created along with a database of high-resolution images. These high-resolution images had 4 times the number of pixels than the low-resolution images. All the images were in grayscale with 8-bit representing a pixel resulting in a total of 256 shades of gray. Low-resolution images were then upscaled to match the resolution of high-resolution images by repeating each pixel 4 times. Then one by one, each of the high-resolution images was assumed to be the correct one, taking its pixels as the target state and the corresponding pixels of the up-scaled image as the initial state. Then following two types of fixed-point search algorithms are performed on them separately.

 \begin{itemize}
    \item One-qubit operation: Performing one iteration of the fixed-point algorithm on a single qubit
representing the individual pixel of the upscaled image. This method is inefficient since, for a low-resolution image with a $p$ number of pixels, we will need to perform $4p$ number of operations. However, this results in the final image that is closer to the high-resolution image. 
\item Four-qubit operation$\colon$ Performing one iteration of the fixed-point algorithm on four qubits representing a 2x2 block of pixels of the upscaled image. This method is more efficient since, for a low-resolution image with a $p$ number of pixels, we will need to perform $p$ number of operations. However, this results in the final image that is closer to the upscaled image.
\end{itemize}

These operations were performed using CQASM simulator \cite{cqasm}, with 256 shots for each measurement. For this particular scenario, where we are increasing the number of pixels in length and width of the image by a factor of 2, we get 4 pixels in the high-resolution image for the corresponding 1 pixel in the low-resolution image. To generalize this, if we increase the number of pixels in the length and width of the image by a factor of $n$, we will get $n^{2}$ times the original number of pixels to preserve the aspect ratio. 
The 4-qubit operation can be generalized as $n^{2}$-qubit operation. If the low-resolution image has a $p$-number of pixels, we can also perform $p \times n^{2}$-qubit operation multiple times recursively to get the desired result. However, this is not possible in the noisy intermediate-scale quantum (NISQ) era  of quantum computing. This is why two types of operations, namely, the one-qubit and the $n^{2}$-qubit operations are used to represent the two different numbers of recursions. For the one-qubit operation, the number of times the fixed-point search will be used is $p \times n^{2}$, whereas, for the $n^2$-qubit operation, it will be used $p$-times. After storing the final image for each corresponding high-resolution image, they were input into the MATLAB OCR to retrieve the confidence value for the corresponding character. The character which yields the highest confidence value is assumed to be the correct character. For a classical computer, let $m$ be the number of bits being used to represent a pixel. If we want
to scale up the low-resolution image by a factor of $n^{2}$ and the $n^{2}$-qubit operation can be used once
or it can be used recursively then we can compare the time complexity required to apply this transformation
once. To apply the same transformation, classical computer will take $\mathcal{O}(m(2n^{2})^3
(n^{2}2n^{2}+\log(m)))$\footnote[1]{This is bit complexity for the schoolbook method of the matrix which has the complexity of $\mathcal{O}(n^{3})$  for the multiplication of two $n \times n$ matrices. The RAM model for calculating complexity has not been used because of the dependence on number of bits $m$ used for each element. It was proved by Strassen that there exist algorithms that are better \cite{alman2021refined}. Currently, the best-known matrix multiplication algorithm has been shown to have the complexity $\mathcal{O}(n^{2.3728596})$ \cite{strassen1969gaussian} for matrix multiplication of two
$ n \times n$ matrices.}. For the proposed classical-quantum hybrid algorithm, we will need to input the unitary matrices
for fixed-point Search. This will require $\mathcal{O}(2^{m})$ time since multiple measurements will need to
be made to assign the pixel value in the processed image. In most cases, $m = 8$, so the proposed
algorithm is mostly faster than the classical counterpart, particularly for large values of $n$.
This algorithm will be most efficient if the fixed-point  algorithm is performed recursively on the
entire image. For a low-resolution image with $p$ number of pixels, the complexity for one recursion
will be $\mathcal{O}(p2^{m})$ \footnote[2]{Here, the additional p has been added to preserve the calculated mean of ”Match Percent”, in Results} for the quantum computer. For classical computation, the complexity will be $\mathcal{O}(m(2^{pn^{2}}))^{3}
(pn^{2}2^{pn^{2}}+ \log(m))$. This means that with each recursion, the quantum algorithm will be much faster.

\begin{figure}[htbp]
    \includegraphics[width=7cm, trim=0cm 0cm 0cm 0cm]{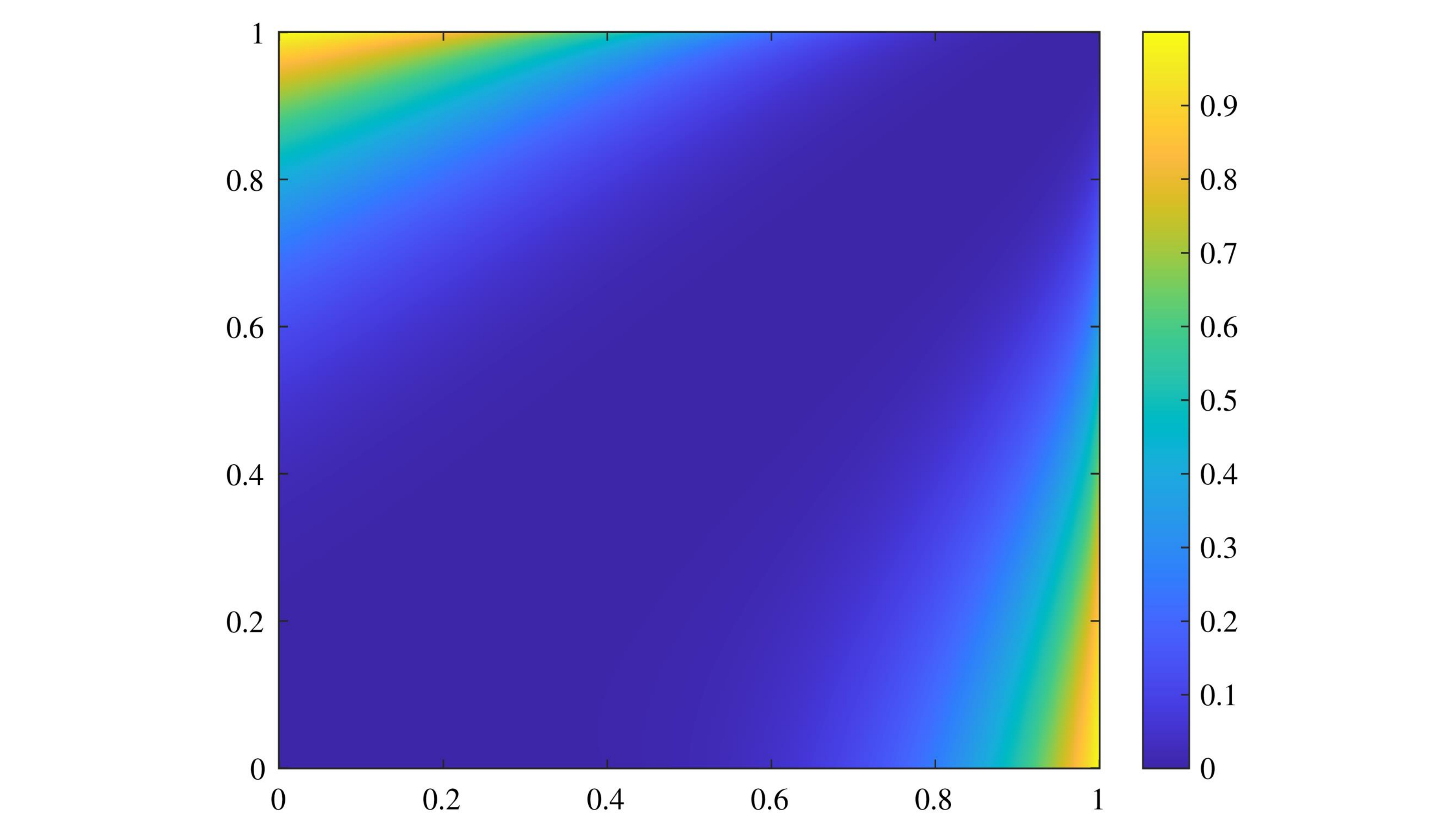}
    \caption{(Color online) Similar to Fig. \ref{fig6Improvement}, here too, the $x$-axis is the probability of measuring the $\ket{0}$ state for
the initial state $\ket{s}$ and the $y$-axis is the probability of measuring the $\ket{0}$ state for the target state $\ket{w}$.
The color map represents the difference between the probability of measuring the $\ket{0}$ state for the final
state and the probability of measuring the $\ket{0}$ state for the target state $\ket{w}$. We can see that as we
move away from the $45^{\rm{o}}$ line, the difference between the final state and the target state $\ket{t}$ increases
monotonically.}\label{fig7}
\end{figure}

\begin{widetext}
\begin{center}
    \begin{table}[h]
\begin{tabular}{|c|c|c|c|}
\hline
 Character & Confidence Value (LRI) & 1 qubit & 4-qubit\\
     \hline
0 & X & 0.92174 & 0.877482 \\
\hline
 1 & 0.892 & 0.925666 & 0.929506 \\
\hline
2 &0.8271 &0.90625 &0.888177 \\
\hline
 3 & 0.8072 &0.90136 &0.884857\\
\hline
4 &0.8279 &0.85005 &0.8472 \\
\hline
5 & 0.8564 & 0.898562 & 0.891324 \\
\hline
6 & 0.857 & 0.912485 & 0.8764556 \\
\hline
7 & X & 0.939333 & 0.934945 \\
\hline
8 & X & 0.86497 & 0.809763 \\
\hline
9 & 0.8072 & 0.874111 & 0.844882 \\
\hline
A & 0.7018 & 0.8341 & 0.831342 \\
\hline
B & 0.8991 & 0.91372 & 0.902215 \\
\hline
C & 0.8639 & 0.958511 & 0.906628 \\
\hline
D & 0.9285 & 0.932506 & 0.95316 \\
\hline
E & 0.9351 & 0.924665 & 0.911977 \\
\hline
F & 0.9497 & 0.938809 & 0.920079 \\
\hline
G & 0.9018 & 0.941417 & 0.905742 \\
\hline
H &0.9134 &0.971099 &0.979016 \\
\hline
I &0.9337 &0.929127 &0.922976 \\
\hline
J &0.9257 &0.937303 &0.888911 \\
\hline
K &0.896 &0.924054 &0.925211 \\
\hline
L &0.9959 &0.995234 &0.99573 \\
\hline
M & 0.8926 &0.945714 &0.892059 \\
\hline
N &0.9001 &0.955811 &0.977005 \\
\hline
O &X &0.929338 &0.896024 \\
\hline
P &0.8889 &0.911954 &0.928999 \\
\hline
Q &0.7479 &0.917528 &0.843384 \\
\hline
R &0.8974 &0.916192 &0.892154 \\
\hline
S &0.825 &0.891374 &0.909354 \\
\hline
T &0.9452 &0.944083 &0.954243 \\
\hline
U &0.939 &0.92857 &0.978524 \\
\hline
V &0.8146 &0.975452 &0.935945 \\
\hline
W &0.8435 &0.957661 &0.914874 \\
\hline
X &0.7127 &0.899264 &0.846405 \\
\hline
Y &0.8049 &0.872792 &0.746638 \\
\hline 
 Z &0.8772 &0.938702 &0.944462 \\
 \hline
\end{tabular}

\caption{The table shows the confidence value for all alphanumeric characters, for the low-resolution image, and after the application of the hybrid algorithm. The first column shows all alphanumeric characters, the second column shows the confidence value of the low-resolution images, and the third column shows the confidence value after the application of our 1-qubit hybrid algorithm. The fourth column shows the confidence value after the application of our 4-qubit hybrid algorithm}\label{confidencevalues}
\end{table}
\end{center}
\end{widetext}
In Table \ref{confidencevalues}, three characters ('0', '7', 'O') shown by X in the second column. These are the  characters which are misidentified by OCR. Interestingly these characters can be identified after processing using our algorithm, both by the 1-qubit algorithm and 4-qubit algorithm. Our 1-qubit algorithm misidentifies character 8 but the 4-qubit algorithm correctly identifies the character. Interestingly, for character 6, algorithms (both 1-qubit and 4-qubit) are found to misidentify the character while OCR correctly identifies '6' in the low-resolution image. For 6, the confidence value of E in the high-resolution image gives a very high confidence value. This bias for character 'E'  for returning high confidence value is evident in Table \ref{cv2} and in \ref{cv3}. The Table \ref{cv3}  shows that the average confidence value of ’0’ when high-resolution image of ’0’ was taken as the target image for rest of the characters is considerably lower than that of ’E’. This can either be due to a biased OCR, or it can be that the shape of ’E’ matches closely with the shape
of characters in rest of the alphabet. The latter theory can be tested if we consider the confidence
value of the rest of the characters when their high-resolution image was taken as the target state and
low-resolution image of ’0’ and ’E’ were processed. If the shape of the letter ’E’ really matches
closely with the shape of the rest of the characters, the OCR should return a high confidence value on
average. However, it can be seen from Table \ref{cv4} that the average confidence value for ’0’ is higher. Therefore, it seems that this algorithm’s results can be improved by training and using an OCR
that is specially trained on images of printed individual characters. It is also important to note that
the typefaces used for printing number plates is usually known. This means we can further train our
OCR for that particular typeface which should provide additional improvements. This is discussed
more elaborately in the last paragraph of \ref{concl}.
However, there is an easier way to find out the results by this algorithm in an ideal case.
It can be done by using the high-resolution image themselves as reference images. By doing this,
we can create a very simple model for pixel-by-pixel image matching. Since the images are in
grayscale, we can calculate the following: 

\begin{equation}
    f(x,y) = 1 - \frac{\sigma_{i}\norm{x_{i}-y_{i}}}{N}
\end{equation}
Here, $x$ and $y$ represent the two images with $N$-number of pixels each and $x_{i}$ and $y_{i}$ represent the
value of the $i^{th}$ pixel of respective images. The calculated value $f(x, y)$ will be referred to as the
’Match Percent’ between the two images. It is important to note that this is a very simplistic model
which can only measure the relative similarity between two corresponding pixels of an image. It can
not replace OCR completely because there may be noise, which can lower the match percent for the
correct character or there may be a case of non-feature matching, which may increase the match
percent of incorrect characters.
\begin{widetext}
\begin{center}
    \begin{table}[h]
\begin{tabular}{|c|c|c|}
\hline
 Character & Confidence Value (0) & Confidence Value (E)\\
     \hline
0 & 0.92174 & 0.859433 \\
\hline
 1 & 0.029358 & 0.764754  \\
\hline
2 &0.029358 &0.800612  \\
\hline
 3 & 0.030201 &0.789907\\
\hline
4 &0.719694 &0.715601  \\
\hline
5 & 0.075472 & 0.872448\\
\hline
6 & 0.069231 & 0.918335  \\
\hline
7 & 0.058823 & 0.826405  \\
\hline
8 & 0.84667 & 0.896808 \\
\hline
9 & 0.056582 & 0.809741  \\
\hline
A & 0.039724 & 0.80018  \\
\hline
B & 0.072253 & 0.800814 \\
\hline
C & 0.110769 & 0.939481 \\
\hline
D & 0.792613 & 0.770146 \\
\hline
E & 0.9351 & 0.924665 \\
\hline
F & 0.9497 & 0.938809 \\
\hline
G & 0.726062 & 0.941417\\
\hline
H &0.035011 &0.954693 \\
\hline
I &0.039724 & 0.921477 \\
\hline
J &0.039724 &0.771872 \\
\hline
K &0.048764 &0.79392 \\
\hline
L &0 &0.953842 \\
\hline
M & 0.047059 &0.806853 \\
\hline
N &0.058823 &0.811308 \\
\hline
O &0.716013 &0.811292 \\
\hline
P &0.082569 &0.864928 \\
\hline
Q &0.62479 &0.716127 \\
\hline
R &0.056582 &0.809852 \\
\hline
S & 0.069231 & 0.810731 \\
\hline
T &0.026311 &0.772327 \\
\hline
U &0.047864 &0.928568 \\
\hline
V &0.722633 &0.746693\\
\hline
W &0.64834 &0.660381\\
\hline
X &0 & 0.752046\\
\hline
Y &0.037257 &0.774703\\
\hline 
Z &0.03121 & 0.881578\\
\hline
Mean &0.224609& 0.8276320\\
 \hline
\end{tabular}

\caption{The table shows the confidence value for all alphanumeric characters, for the low-resolution image, and after the application of the hybrid algorithm. The first column shows all alphanumeric characters, the second column shows the confidence value of the low-resolution images, and the third column shows the confidence value after the application of our 1-qubit hybrid algorithm. The fourth column shows the confidence value after the application of our 4-qubit hybrid algorithm}\label{cv2}
\end{table}
\end{center}
\end{widetext}
\begin{widetext}
\begin{center}
    \begin{table}[h]
\begin{tabular}{|c|c|c|}
\hline
 Character & Confidence Value (0) & Confidence Value (E)\\
     \hline
0 &0.921740 &0.075472 \\
\hline
1 &0.0487640 &0.063717  \\
\hline
2 &0.752476& 0\\
\hline
3 &0.768959 &0.660494\\
\hline
4 &0.583875 &0\\
\hline
5 &0.817715 &0.774370\\
\hline
6 &0.842950& 0.829795\\
\hline
7 &0.066390 &0\\
\hline
8 &0.750651& 0\\
\hline
9 &0.812266 &0.066390\\
\hline
A &0 &0\\
\hline
B &0.863754 &0.881197\\
\hline
C &0.032000& 0\\
\hline
D &0.814780 & 0.105186\\
\hline
E &0.859433 &0.924665\\
\hline
F &0.052499 &0.100000\\
\hline
G &0.707981 &0.664239\\
\hline
H &0.760396 &0.790886\\
\hline
I &0.802360 &0.042440\\
\hline
J &0.123288 &0.056582\\
\hline
K& 0.714194 &0.739672\\
\hline
L &0.850833 &0.0722533\\
\hline
M &0.712009 &0.25623\\
\hline
N &0.647434 & 0\\
\hline
O &0.706801 & 0.024324\\
\hline
P& 0.765699 & 0.706166\\
\hline
Q &0.675175 & 0.591882\\
\hline
R &0.775897 & 0.68769\\
\hline
S &0.755298 & 0.726142\\
\hline
T &0.116788 & 0.024961\\
\hline
U &0.770081 & 0\\
\hline
V& 0.738190 & 0.598832\\
\hline
W &0.622550 & 0.540786\\
\hline
X &0.650041 &0\\
\hline
Y &0.032000 &0.030201\\
\hline
Z &0.788975 &0\\
\hline
Mean &0.602896 &0.306516\\
\hline
\end{tabular}
\caption{The ’Character’ column represents the character in the high-resolution image taken as
the target image and the ’0’ and ’E’ columns represent the confidence value of the corresponding
characters when a low-resolution image of ’0’ and ’E’ is processed respectively.}\label{cv3}
\end{table}
\end{center}
\end{widetext}

\begin{widetext}
\begin{center}
\begin{table}[h]
\begin{tabular}{|c|c|c|c|}
\hline
Character & Match Percent & Mean & Mean Deviation\\
\hline
0 & 0.9848823529 & 0.924672658 & 0.02420893851\\
\hline
1 & 0.9968333333 & 0.8929891068 & 0.02881094166\\
\hline
2 & 0.992127451 & 0.9070144336 & 0.02710763132\\
\hline
3 & 0.9916666667 & 0.9080716231 & 0.03007439179\\
\hline
4 & 0.9917647059 & 0.8939926471 & 0.02426116558\\
\hline
5 & 0.9910588235 & 0.9190220588 & 0.02455718954\\
\hline
6 & 0.9880098039 & 0.9202140523 & 0.02480591866\\
\hline
7 & 0.9912254902 & 0.9054283769 & 0.02753622004\\
\hline
8 & 0.9848627451 & 0.9244586057 & 0.02539584846\\
\hline
9 & 0.9859901961 &0.9152706972 &0.02427874607\\
\hline
A &0.9524411765 &0.8895054466 &0.02356463326\\
\hline
B & 0.9922941176 & 0.9123374183 & 0.02813784495\\
\hline
C &0.9896764706 &0.9185114379& 0.02348202614\\
\hline
D &0.9918431373 &0.8994360022& 0.0297573832\\
\hline
E &0.9954901961 &0.9153924292 &0.02665601549\\
\hline
F &0.9953333333 &0.9167393791 &0.02400290487\\
\hline
G &0.9881960784 & 0.901665305 & 0.02562212539\\
\hline
H &0.9873137255 &0.9031555011 &0.02600571895\\
\hline
I &0.9971470588 &0.9090705338 &0.02148872852\\
\hline
J & 0.9965 & 0.8803474946 & 0.03281148632\\
\hline
K &0.9907058824 &0.906995915 &0.01916344408\\
\hline
L &0.9973235294 &0.9120277778 &0.02217102397\\
\hline
M& 0.9815 &0.8892592593 &0.02196514161\\
\hline
N &0.9786666667 &0.9055234205 &0.02486504478\\
\hline
O &0.9865 &0.9012184096 &0.02780694747\\
\hline
P &0.9924803922 &0.9121176471 &0.0215664488\\
\hline
Q &0.9828431373 &0.8884471678 &0.02911374365\\
\hline
R &0.9899313725 &0.9105966776 &0.02211729908\\
\hline
S &0.9872941176 &0.9196775599 &0.02380440571\\
\hline
T &0.9967745098 &0.8862086057 &0.03107764464\\
\hline
U &0.9417058824 &0.907370915 &0.02780455701\\
\hline
V & 0.988 &0.9003551198 &0.01648517308\\
\hline
W& 0.9768627451 &0.876333878 &0.01377795933\\
\hline
X &0.9875784314& 0.8996160131 &0.02496205519\\
\hline
Y & 0.9922058824& 0.898956427 &0.0262456427\\
\hline
Z &0.9909705882 &0.9096955338 &0.02583717623\\
\hline
Mean &0.9873888889 &0.9050470982 &0.02503693239\\
 \hline
\end{tabular}
\caption{1-qubit operation: The ’Character’ column represents the character in the low-resolution image. The ’Match Percent’ column gives the calculated value of the match percent for the processed image when the high-resolution image (target image) contains the same character as in the low-resolution
image. The ’Mean’ column gives the mean match percent for all the characters in the alphabet. The ’Mean Deviation’ columns give the mean of the magnitude of absolute deviation from the mean.}\label{cv4}
\end{table}
\end{center}
\end{widetext}

\section{Conclusion} \label{concl} We have devised a hybrid algorithm for the identification of the missing information in a low-resolution image of the alphanumeric character. Our algorithm could identify the missing information in the low-resolution image using the high-resolution images of the alphanumeric character set. The confidence value for all known low-resolution alphanumeric characters is shown in the second column of Table \ref{confidencevalues}. In the third column, we chose the maximum confidence value out of all the values corresponding to high-resolution images, for all known alphanumeric characters. It is clearly visible from the confidence values shown in the third column that the algorithm is able to identify the known low-resolution image of the alphanumeric character correctly. The confidence value for the 4-qubit version is shown in the last column of Table \ref{confidencevalues}. One can clearly see the increased confidence value after the application of the 4-qubit hybrid algorithm. Though this scheme is restricted to typeset characters. It should also be possible to identify the missing information in handwritten characters by combining the deep learning algorithms with our algorithm. 

Though this work has mostly studied the case of English alphanumeric character set for a particular
typeface, it can be extended to other natural languages and to different typefaces. However, it is
important to note that natural languages have not evolved to take into account how well an OCR can
be trained to identify its characters. This poses a very interesting question about the fundamental
lower limit of visual information needed to tell each character in alphabet apart. Ultimately, an
image is just information that needs to be processed and the amount of information depends on the
resolution of the image. There must be a limit to the number of pixels, below which information
can’t be sufficient as to identify the correct character. This limit will be different for different
natural languages and for different typefaces. This algorithm can only work in a case when there
is sufficient information, so it will be important to know where the limit is, beyond which we can’t
trust this algorithm. Thus, we can also find the best typeface to use to print something in a practical situation where there is a possibility of involving OCR algorithm to identify a low-resolution character.
Another thing that is important is that the OCR we are relying on to give the confidence value
must have few properties $\colon$
\begin{itemize}
\item The OCR must be trained on the data-set of printed text, particularly on the typeface used to
print such texts.
\item It should not use Language Techniques, Dictionary Look-ups, or any other contextual editing
or correction technique. For special cases, it may use these, but not when used for something
like automatic number-plate recognition.
\item It should perform pre-processing techniques before resolution enhancement, not after.
To train such an OCR would require a lot of resources like huge data-sets of text printed in same
typeface and a suitable training algorithm.
\end{itemize}

In this algorithm, the number of shots chosen (number of times measurement is done) has been
taken as simply $2m$, this is because it is the least number of measurements required to get the
value required to be assigned by $m$ number of classical bits. Although, this has produced desirable
results, because of the probabilistic nature of measurement, it may not be enough to ensure accuracy
for larger values of scale up factor $n$. This value will also depend on decoherence or another form of noise. Therefore, some form of quantum error correction may also be needed. Lastly, the ideal number of recursions required will need evaluation. The fixed-point algorithm
will almost always converge to the target state with each recursion. This means that if an excessive
number of recursions are used, only the pixels which are either completely black or completely
white in the low-resolution image will remain different. Other than that pixel, the processed image
will be the same as the high-resolution image taken from the database.  
\section*{Acknowledgement}  A Pal and A Pathak acknowledge support from the QUEST scheme of the Interdisciplinary Cyber-Physical Systems (ICPS) program of the Department of Science and Technology (DST), India, Grant No.: DST/ICPS/QuST/Theme-1/2019/14 (Q80). A Pal also thanks, Indian Institute of Science Education and Research, Kolkata for the support. A Shukla thanks to Interuniversity microelectronics centre, Belgium and University of Hasselt, Diepenbeek, Belgium for the financial support and computational facilities. All authors thank Kishore Thapliyal for his interest and useful technical feedback on this work. A Pal thanks Chiranjib Mitra for his interest in this work.
\bibliography{ref1}
\end{document}